\documentclass{article}

\usepackage{arxiv}

\usepackage[utf8]{inputenc} 
\usepackage[T1]{fontenc}    
\usepackage{hyperref}       
\usepackage{url}            
\usepackage{booktabs}       
\usepackage{amsfonts}       
\usepackage{nicefrac}       
\usepackage{microtype}      
\usepackage{graphicx} 
\usepackage{doi}
\usepackage{float}
\usepackage{multirow} 
\usepackage{amsmath}
\usepackage{stmaryrd}
\usepackage{tabularx}
\usepackage{amssymb} 
\usepackage{tabularx, microtype, booktabs, multirow}
\usepackage{subcaption}
\usepackage{comment}
\usepackage{pifont}
\makeatletter
\usepackage[numbers]{natbib}

\usepackage[toc,page]{appendix}

\usepackage{xcolor}

\title{Enhancing low dose Computed Tomography Images using Consistency Training Techniques}

\author{ \href{https://orcid.org/0000-0000-0000-0000}{\includegraphics[scale=0.06]{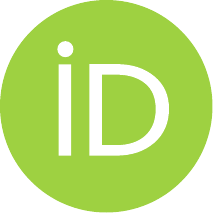}\hspace{1mm}Mahmut S. Gokmen} \\
	Department of Computer Science\\
	University of Kentucky\\ 
	\texttt{mselmangokmen@uky.edu} \\
	\And
	\href{https://orcid.org/0000-0000-0000-0000}{\includegraphics[scale=0.06]{orcid.pdf}\hspace{1mm}Cody Bumgardner} \\
	Department of Internal Medicine\\
        Institute for Biomedical Informatics\\
	University of Kentucky\\  
	\texttt{cody@uky.edu} \\
  \And
	\href{https://orcid.org/0000-0000-0000-0000}{\includegraphics[scale=0.06]{orcid.pdf}\hspace{1mm}Jie Zhang} \\
	Department of Radiology\\
	University of Kentucky\\  
	\texttt{jie.zhang1@uky.edu} \\
 \And
 \href{https://orcid.org/0000-0000-0000-0000}{\includegraphics[scale=0.06]{orcid.pdf}\hspace{1mm}Ge Wang} \\
        Department of Biomedical Engineering\\
	Rensselaer Polytechnic Institute\\  
	\texttt{wangg6@rpi.edu} \\
 \And
	\href{https://orcid.org/0000-0000-0000-0000}{\includegraphics[scale=0.06]{orcid.pdf}\hspace{1mm}Jin Chen} \\
	Department of Medicine\\
        Department of Biomedical Informatics and Data Science\\
	University of Alabama at Birmingham\\  
	\texttt{jinchen@uab.edu} \\
}

\renewcommand{\shorttitle}{ENHANCING LOW DOSE COMPUTED TOMOGRAPHY IMAGES
USING CONSISTENCY TRAINING TECHNIQUES}

\begin{document}
\maketitle

\begin{abstract} 
Diffusion models have significant impact on wide range of generative tasks, especially on image inpainting and restoration. Although the improvements on aiming for decreasing number of function evaluations (NFE), the iterative results are still computationally expensive. Consistency models are as a new family of generative models, enable single-step sampling of high quality data without the need for adversarial training. In this paper, we introduce the beta noise distribution, which provides flexibility in adjusting noise levels. This is combined with a sinusoidal curriculum that enhances the learning of the trajectory between the noise distribution and the posterior distribution of interest, allowing High Noise Improved Consistency Training (HN-iCT) to be trained in a supervised fashion. Additionally, High Noise Improved Consistency Training with Image Condition (HN-iCT-CN) architecture is introduced, enables to take Low Dose images as a condition for extracting significant features by Weighted Attention Gates (WAG).Our results indicate that unconditional image generation using HN-iCT significantly outperforms basic CT and iCT training techniques with NFE=1 on the CIFAR10 and CelebA datasets. Moreover, our image-conditioned model demonstrates exceptional performance in enhancing low-dose (LD) CT scans.

\end{abstract}


\section{Introduction}

X-ray computed tomography (CT) is essential in both diagnosis and treatment, with applications ranging from detecting internal injuries and tumors to surgical planning.To minimize the harmful effects of low-dose ionizing radiation, many studies focus on achieving high-quality denoising while keeping the dose as low as reasonably possible.
Recent studies reveals that generative tasks has a remarkable success to increase quality of low dose CT scans. The most commonly used techniques due to their ease of application are Non-local Means (NLM) and Block-Matching 3D (BM3D), both of which can enhance low-dose CT (LDCT) performance. However, despite their utility, these post-processing methods often fall short of meeting clinical requirements \cite{mean_filter, BM3D}.

Recent advancements in deep learning showed that generative models are highly capable of meeting clinical requirements for LDCT denoising~\cite{DDPM_LDCT, DDPM, DDPM_Survey, GAN_survey}. Especially, Diffusion Probabilistic Models (DDPM) models outperform other techniques, especially GANs, upon the task of image denoising by iteratively recovering data~\cite{DDPM_vs_GAN}. The core working principle of DDPM relies on recovering gradually perturbed data from a Gaussian distribution to the original data distribution. This iterative process has a strong capacity to predict and restore missing or noisy pixels by replacing them with the most likely candidates according to the data distribution. 
This iterative denoising technique ensures that DDPM models remain generative throughout the process, gradually converging to the original data distribution with high accuracy.

The primary challenge with DDPM models lies in the iterative time-consuming denoising process. While recent studies have reduced the number of function evaluations (NFE) to as few as 20 steps, generating the highest quality samples typically requires at least 80 NFEs. To address this issue, consistency models (CM) have recently emerged as a family of generative models, utilizing rapidly evolving knowledge distillation training techniques. 
Unlike DDPM models, such as score-based diffusion models~\cite{score_based}, CM does not require multiple sampling steps to produce high-quality samples; it can generate results in a single step. Also, CMs preserves flexibility and computation theory for generating samples in different number of inference steps. 

Consistency models uses two different training techniques, which are represented as consistency distillation (CD) and consistency training (CT)~\cite{Song2023-cm}. The CD training requires a pre-trained model for distilling the knowledge into consistency model. In contrast, the CT training allows the consistency model to learn directly from data. Although recent studies shows high-performance consistency models successfully built with the CD training, CD relies on pre-trained models and requires high computational resource for distilling the data from pre-trained models. Additionally, model performance is limited by the pre-trained models. Song et al~\cite{Song2023-imp} demonstrated that the CT training has the potential to outperform the CD training by eliminating Exponential Moving Average (EMA) for the teacher network and improving curriculum with a step-wise increase relied on the current training steps. Furthermore, noise scheduling could also be improved by adopting the log-normal distribution to sample noise levels. 

While full of potential, the central problem is the CT training is  its sensitivity to noise distribution during training. Although the log-normal noise distribution heavily weights low noise levels in mini-batches, our experiments on various noise distributions reveal that a more balanced noise distribution, spanning from the lowest to the highest noise levels, produces better results. Another issue with consistency models is the performance loss that arises from zero-shot editing. Unlike multi-step denoising, consistency models use ODE solvers to reach the original data distribution in a single step. This approach leaves no room for trade-offs between computation and image quality, resulting in zero tolerance for any degradation. Additionally, the high computational demands of processing medical images, which often have high resolutions, necessitate the design of a new model. It is also important to provide high quality denoised samples in a short of training period of the model.

To overcome the problems arise from log-normal distribution, it is introduced beta scheduling which has two parameters $\alpha$ and $\beta$ provides flexibility for adjusting the weights of noise levels in a mini-batch. This beta noise scheduling significantly enhances unconditional image generation while requiring only half the number of parameters compared to previous models.

Another challenge lies in the curriculum, which determines the variety of noise levels encountered during model training. The latest iCT approach \cite{Song2023-imp} utilizes an improved curriculum technique that doubles the variety of noise levels presented to the model every 50k training steps. While this improved curriculum is effective in representing various noise levels and preventing overfitting, it can reduce the traceability of trajectory points due to marginal changes in the curriculum. To address this, we propose a curriculum based on a sinusoidal function, which gradually decreases the rate at which the number of noise varieties increases across levels, thereby maintaining the traceability of trajectory points.

\begin{figure}[h]
\centering
\includegraphics[width=0.8\linewidth]{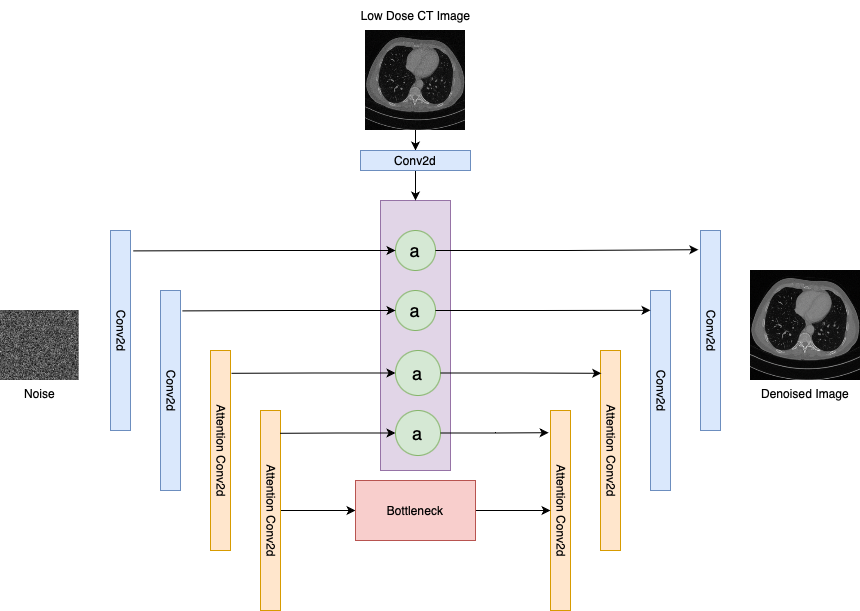}
\caption{The illustration of image conditioned consistency model   }
\label{fig:model_arch}
\end{figure}

Additionally, the architecture proposed in this study for medical image denoising can effectively extract common features between an unconditionally generated image and a conditionally represented image, as illustrated in Figure \ref{fig:model_arch}. With this technique, instead of adding noise to an LDCT image, we allow the model to generate an unconditional medical image while matching common features with the LDCT image, which is used as a condition. Thus, the model has a able to generate an unconditional image within under supervision of LDCT image. To mitigate the dominance of the LDCT-conditioned image and prevent additional computational costs, the weighted attention gate modules have been introduced in this study.

Our contributions address key challenges in consistency models by enhancing training techniques and proposing a novel architecture for medical image denoising. These improvements offer more efficient and effective solutions compared to existing methods.

\begin{itemize}
    \item Introduced \textbf{beta noise scheduling} for improved unconditional image generation with reduced parameters.
    \item Proposed a \textbf{sinusoidal curriculum} to maintain trajectory point traceability by gradually adjusting noise level increments.
    \item Developed a new architecture for \textbf{medical image denoising} using weighted attention gate modules for balanced supervision. 
    
\end{itemize}

\section{Background}
\subsection{Consistency Models}

Consistency models \cite{Song2023-cm}, relies on continuous-time definition of diffusion models which generate data by gradually transforming the data distribution,$p_{\text{data}}$, into a noise distribution using stochastic differential equation (SDE). These models then learn to reverse this process to generate data from noise. Notably, the SDE has a related ordinary differential equation (ODE) known as the probability flow (PF) ODE \cite{score_based}.According to definition of  score based generative models, PF ODE shapes into

\begin{equation}
dx = -\sigma \nabla_x \log p_{\sigma}(x) \, dt, \sigma \in [\sigma_{min},\sigma_{max}]
\end{equation}
for $ \nabla_x \log p_{\sigma}(x)$ is score function for perturbed data distribution $p_{\sigma}(x) \approx  p_{\sigma_{min}}(x)$. $\sigma_{min}$ is defined as $0.002$ to maintain stability and $\sigma_{max}=80$ is chosen as reasonable by \cite{Karras} to provide $p_{\sigma}(x) \approx \mathcal{N}(0, \sigma_{\text{max}}^2 \mathbf{I})$. Solving the probability flow ODE from $\sigma(t+1)$ to $\sigma(t)$ allows the model to transform a sample $x_{\sigma_{t+1}} \approx p_{\sigma_{t+1}}$ to  $x_{\sigma_{t}} \approx p_{\sigma_{t}}$.  The provided bijective mapping between $x_{\sigma} \approx p_{\sigma}(x)$ and $x_{\sigma_{min}} \sim p_{\sigma_{min}}(x) \approx p_{data}(x)$ maintains $consistency$ and which is denoted as $f^* : (\mathbf{x}_{\sigma}, \sigma) \mapsto \mathbf{x}_{\sigma_{\text{min}}}$. 

A consistency model which denoted as $f_{\theta}(\mathbf{x}, \sigma)$ is parameterized to meet boundary condition and transforming it into a differentiable form, it is parameterized as it is defined in \cite{Song2023-cm}. 

\begin{equation}
f_{\theta}(\mathbf{x}, \sigma) = c_{\text{skip}}(\sigma) \mathbf{x} + c_{\text{out}}(\sigma) F_{\theta}(\mathbf{x}, \sigma),
\end{equation}
Where $F_{\theta}(\mathbf{x}, \sigma)$ is a free-from network. To train consistency models, probability flow ODE is discretized using noise sequences are ranging from $\sigma_{min}$ to $\sigma_{max} = \sigma_N$. Discretization of these noise sequences is denoted as $\sigma_i = \left( \sigma_{\text{min}}^{1/\rho} + \frac{i - 1}{N - 1} \left( \sigma_{\text{max}}^{1/\rho} - \sigma_{\text{min}}^{1/\rho} \right) \right)^{\rho}$ for $i \in \llbracket 1, N \rrbracket$, $\rho=7$ in \cite{Karras}.

Consistency models are trained using the consistency matching loss
\begin{equation}
\mathbb{E} \left[ \lambda(\sigma_i)d(f_{\theta}(\tilde{\mathbf{x}}_{\sigma_{i+1}}, \sigma_{i+1}), f_{\theta^-}(\tilde{\mathbf{x}}_{\sigma_i}, \sigma_i)) \right], \tag{5}
\end{equation}
where
\begin{equation}
\tilde{\mathbf{x}}_{\sigma_i} = \mathbf{x}_{\sigma_{i+1}} - (\sigma_i - \sigma_{i+1})\sigma_{i+1} \nabla_{\mathbf{x}} \log p_{\sigma_{i+1}}(\mathbf{x}) \big|_{\mathbf{x}=\mathbf{x}_{\sigma_{i+1}}} \tag{6}
\end{equation}
is a single-step in the reverse direction solving the PF ODE in Eq. (1). Here, \(\lambda(\cdot)\) is a positive weighting function, and \(d(\cdot, \cdot)\) is determined as $1$ in \cite{Song2023-cm}. It is  suggested using the Learned Perceptual Image Patch Similarity (LPIPS) as the metric function \cite{Song2023-cm}. The expectation in Eq. (5) is taken with respect to \(i \sim U[1, N-1]\), a uniform distribution over integers \(i = 1, \ldots, N-1\), and \(\mathbf{x}_{\sigma_{i+1}} \sim p_{\sigma_{i+1}}(\mathbf{x})\). The objective in Eq. (5) is minimized via stochastic gradient descent on parameters \(\theta\) while \(\theta^-\) is updated using the exponential moving average (EMA)
\begin{equation}
\theta^- \leftarrow \text{stopgrad}(\mu \theta^- + (1 - \mu) \theta), \tag{7}
\end{equation}
where \(0 \leq \mu \leq 1\) is the decay rate. \(f_{\theta}\) and \(f_{\theta^-}\) are referred to as the student and teacher networks, respectively. 

The training technique for consistency models can be divided into two categories: consistency distillation (CD) and consistency training (CT). The consistency distillation technique relies on utilizing a pre-trained EDM model as a teacher model, denoted as \( f_{\theta^-}(\tilde{\mathbf{x}}_{\sigma_i}, \sigma_i) \).

Consistency distillation (CD) training technique surpasses consistency training (CT) in FID results when compared. It is crucial that the best results for a model trained with CD can only match the best results of the teacher model trained with EDM. To overcome this obstacle, \cite{Song2023-imp} suggests an improved training technique for CT, which generates better FID results.

\subsection{Improved Training Techniques for Consistency Models}
Improved training techniques for CT moves one step forward isolation training for consistency models \cite{Song2023-imp}. The modifications are utilized for improved consistency models (iCM) consist of curriculum, elimination of EMA and teacher model, replacing LPIPS loss function with pseudo huber loss and the noise distribution changed from uniform to lognormal noise distribution.
The first modification, setting $N$ to 1281 in $N(k)$, provides a good balance between bias and variance compared to CM training. Experimental results show that changing $s_0$ from $2$ to $10$ and $s_1$ from $150$ to $1280$ yields the best generative performance for iCT.
\begin{equation}
N(k) = \min(s_0 2^{\left\lfloor \frac{k}{K'} \right\rfloor}, s_1) + 1, \quad K' = \left\lfloor \frac{K}{\log_2 \left[ s_1 / s_0 \right] + 1} \right\rfloor, \text{where } K=1280
\end{equation}
$iCT$ utilizes the same noise scheduling as it is described in \cite{Karras} and Equation \ref{karras_scheduling} which emphasizes high weighted low noise levels and corresponds to $p(\log \sigma) = \sigma \frac{\sigma^{1/\rho - 1}}{\rho (\sigma_{\max}^{1/\rho} - \sigma_{\min}^{1/\rho})} \quad \text{as } N \to \infty.$. 
\begin{equation}
\sigma_i = \left( \sigma_{\min}^{1/\rho} + \frac{i - 1}{N - 1} \left( \sigma_{\max}^{1/\rho} - \sigma_{\min}^{1/\rho} \right) \right)^{\rho} \quad \text{for } i \in \llbracket 1, N \rrbracket, \text{ and } \rho = 7,
\label{karras_scheduling}
\end{equation}

Besides the exponential curriculum and Karras noise scheduling, to emphasize lower noise levels in the noise distribution during training, the modification employed for iCM includes a log-normal noise distribution on image batches, which significantly assigns low weights to high noise levels.

\begin{equation}
p(\sigma_i) \propto \mathrm{erf} \left( \frac{\log(\sigma_{i+1}) - P_{\mathrm{mean}}}{\sqrt{2} P_{\mathrm{std}}} \right) - \mathrm{erf} \left( \frac{\log(\sigma_i) - P_{\mathrm{mean}}}{\sqrt{2} P_{\mathrm{std}}} \right),
\end{equation}
where \( P_{\text{mean}} = -1.1 \) and \( P_{\text{std}} = 2.0 \). This log-normal noise schedule leverages sampling quality and significantly decrease FID scores. 
Addition to lognormal noise distribution, to increase the emphasize on lower noise levels which is provided by lognormal noise distribution, the loss weighting is adjusted as 
\begin{equation}
\lambda(\sigma_i) = \frac{1}{\sigma_{i+1} - \sigma_i}.
\end{equation}
The refined weighting function notably improves sample quality in consistency training by assigning smaller weights to higher noise levels. This approach addresses the issue of the default uniform weighting function, which assigns equal weights to all noise levels and is found to be suboptimal. By reducing the weighting as noise levels increase, the new method ensures that smaller noise levels, which can influence larger ones, are weighted more heavily.

Improved training techniques for CT eliminates teacher model and EMA update during training. The underlying reason for elimination of EMA and teacher model is unbiased trajectory mapping between student and teacher model and the decay rate is updated as $\mu(k)=0$. 
The loss function LPIPS employed for CT is replaced with the pseudo-Huber loss due to undesirable bias in evaluation.
\begin{equation}
d(x, y) = \sqrt{\| x - y \|_2^2 + c^2} - c
\label{eq:pseudo_huber_loss}
\end{equation}

\section{Methods}
\subsection{Architecture}
The proposed approach, High Noise Improved Consistency Training (HN-iCT), shares the same backbone as the consistency models used for image generation. We categorized this approach into two sections: the first focuses on unconditional image generation, which involves modifications to the curriculum and noise distribution, while the second is conditional image generation (HN-iCT-CN), which requires a different architecture in addition to the proposed curriculum and beta noise distribution, as shown in Figure \ref{fig:model_arch}.
 
For unconditional image generation, HN-iCT relies on the same U-Net architecture as implemented for consistency models \cite{Song2023-cm}. For image conditional  generation, the most well-known technique is label embedding or class embedding. Generally, this type of embedding is implemented by summing encoded features irrespective of the dimensions of the input size.  In other cases, such as when an image is given as condition to a model, it is important to choose what type of embedding will be employed while training. In this study, AG (Attention Gate) modules are utilized to extract common spatial features between the image chosen as condition and input image \cite{attention_U_NET}. Further more, AG modules are modified as Weigted Attention Gate (WAG) to prevent incompatibility between other U-NET components and adapted according to HN-iCT model. Figure \ref{fig:model_arch} represents general architecture used for image conditional training, comprehends iAG modules.

\begin{figure}[h]
\centering
\includegraphics[width=0.7\linewidth]{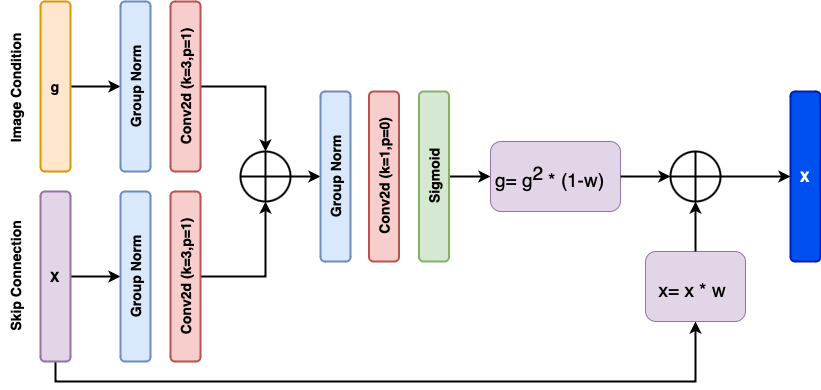}
\caption{The illustration of  WAG unit used in proposed architecture. Weight($w$) parameter is determined as $0.8$.  }
\label{fig:imp_AG}
\end{figure}

\textbf{Weighted Attention Gate (WAG)}: During the encoding process, gradually downsampled features retain essential structural information about the image that will be denoised in the decoding process. While it is possible to reconstruct the denoised image from the latent space, the output may include noisy pixels due to the loss of structural details. To achieve a clean image at the end of the decoding process and preserve structural information, it is necessary to evaluate the skip connections and decoded features. The attention gates are designed to assess both global structural information and pixel-wise details to reconstruct denoised images effectively. In our implementation, the Weighted Attention Gate (WAG) incorporates a learnable weighting mechanism, where the attention map is squared to sharpen common features between the skip connection and the conditioned input. The gate ensures that the skip connections do not dominate the reconstruction while preserving the spatial-temporal features. Additionally, extracted features from the conditioned input are scaled by a weight parameter (defaulting to 0.8) and combined with the skip connection output to achieve a balanced reconstruction. The architecture of  WAG is represented in Figure \ref{fig:imp_AG}.

\subsection{Training Technique}
\textbf{High Noise Level Effect on Performance.}
Noise distribution in mini-batches plays a crucial role in teaching the model the trajectories between noise levels. To provide a high-confidence trajectory, it is important to have a noise distribution that includes a wide variety of noise levels. A noise distribution with a high variety of noise levels increases the number of high noise levels in the distribution, in contrast to the log-normal noise distribution used in iCT training techniques \cite{Song2023-imp}.

In this section, we evaluate the experimental results conducted within the scope of adding extra high noise levels manually to scheduled noises with a log-normal distribution. As it is claimed 'High Noises is a Must', the experiments begin by adding high-noise levels gradually on the mini-batches based on percentage ratio of the mini-batch size. This experiments proves that necessities of a noise scheduling comprising high-level noises beside high weighted low-level noises in a noise scheduling. High noise levels ranging between $40$ and $80$ are randomly added to mini-batches to increase the weight of these high noise levels. The experiments conducted with additional high noise levels reveal that denoising performance significantly increases when the high noise levels constitute up to $4\%$ of the mini-batches. Additional experimental details are provided in appendix \ref{appendix_A}.

\textbf{Beta Distribution.}
The beta distribution is a continuous probability distribution defined on the interval [0, 1], commonly used to model the behavior of random variables that are constrained within this range. It is parameterized by two positive shape parameters, denoted as $\alpha$ (alpha) and $\beta$ (beta), which determine the distribution's shape. The probability density function (PDF) of the beta distribution is given by:

\begin{equation}
f(x; \alpha, \beta) = \frac{x^{\alpha-1}(1-x)^{\beta-1}}{B(\alpha, \beta)} \quad \text{where} \quad B(\alpha, \beta) = \int_0^1 t^{\alpha-1} (1-t)^{\beta-1} \, dt
\end{equation}

for $0 \leq x \leq 1$. The parameters $\alpha$ and $\beta$ influence the skewness and kurtosis of the distribution, allowing for a wide range of shapes including uniform, U-shaped, and J-shaped distributions. This flexibility makes the beta distribution particularly useful in Bayesian statistics, where it is often employed as a prior distribution for probabilities and proportions. \\
The flexibility of the beta distribution makes it highly adaptable for various noise distributions, particularly for those aiming to adjust the weight of noise levels within a mini-batch \cite{beta_diff}. Particularly, it is possible increasing the weight of high level noises in distribution up to $4\%$ by adjusting $\alpha=0.5$ and $\beta=5$. CLAIM and modify: The set of parameters were empirically determined by using CIFAR10.

\textbf{Sinusoidal Curriculum.}
The proposed sinusoidal curriculum, inspired by the sinus function, offers a significant advancement in the training of consistency models by introducing a smooth, continuous progression of the number of timesteps for each noise distribution on mini-batch. Unlike improved curriculum that rely on abrupt or stepwise changes, the sinusoidal curriculum leverages the natural oscillation of the sine function to modulate the variety of noise levels encountered during training. This gradual adjustment ensures that the model experiences a consistent and well-distributed range of noise levels, enhancing the stability and robustness of the learning process.

\begin{equation}
    N(k) = \min\left( \left\lceil \left| \left( s_1 \cdot \sin\left(\frac{\pi \cdot 3 \cdot k}{2 \cdot K}\right) \right) + s_0 \right| \right\rceil + 1, s_1 + 1 \right)
\end{equation}

\begin{equation}
\sigma(i) = \sin\left(\frac{\pi \cdot i}{2 \cdot N}\right) \cdot \Delta t + t_0
\end{equation}

The sinusoidal curriculum is governed by key parameters. The initial timestep $s_0$ sets the starting point, while the difference between final number of time steps and initial the number of time steps \( \Delta t = s_1 - s_0 \) controls the amplitude of the sinusoidal curve. The total number of time steps \( N \) defines the schedule length, with the sine function scaled by \( \pi \) to ensure a smooth, gradual adjustment across timesteps. To prevent marginal increases, $s_0$ and $s_1$ are set to 20 and 250, respectively, providing broad yet stable coverage of noise levels and enhancing training stability.
\section{Experiments}

\subsection{Experimental Setup} 
\textbf{Datasets.} For unconditional image generation, we consider CIFAR10 \cite{cifar10}, which includes 50K training images; CelebA 64x64 \cite{celeba}, which contains 162,770 training images of individual human faces; and Butterflies \cite{butterfly} 256x256 dataset, which provides 1K images with different species of butterflies. 

For image conditional learning, the dataset sourced from the Mayo Clinic, as used in the AAPM low-dose CT grand challenge, was utilized \cite{LDCTdataset}. The data is reconstructed on a 512x512 pixel grid with a slice thickness of 1 mm and a medium (B30) reconstruction kernel. The first eight patients provided training data, resulting in a total of 4800 slices, while the remaining two patients were used for validation, contributing a total of 1136 slices. This demonstrates the model’s effectiveness on medical data and validates that the proposed training technique yields good results across different datasets.  

\textbf{Image Conditioning.} The image conditioning technique employed in our model takes Low dose  512x512 reconstructed CT slices as a condition for denoising process. No additional pre-processing is applied on low dose slices before submitting it as a condition. It is a similar approach such as label embedding. 

\textbf{Model Configurations.} It is identified several configurations for HN-iCT model. This configurations depends on capacity of the hardware for training the model. Additionally, the configurations for other models utilized in experiments  are provided in Table \ref{tab:model_conf}. To make a fair comparison, all hyper parameters such as optimizer type, learning rate and dropout kept as same with HN-iCT model except the number of residual blocks which depends on hardware limitations.
The smallest model configuration of HN-iCT is utilized in our experiments to prove stability and robustness of our approach. Other options of the HN-iCT model were not used due to hardware limitations and extended training time, but these options are available in our public repository for researchers who wish to experiment further.
\begin{table}[hbt!]
\centering
\begin{tabular}{lcccccc}
\toprule
\textbf{Model Name} & \textbf{\#Number of Res. Blocks} & \textbf{Base Channel Size}    & \textbf{\#Params} \\
\midrule
\multicolumn{4}{l}{\textbf{Unconditional Model}} \\
HN-iCT-Small & 2 & 128   & 234M \\
HN-iCT-Small/H (High Res.) & 2 &  192   & 527M \\
HN-iCT-Mid & 4 & 128    & 373M \\
HN-iCT-Large & 6 & 128    & 512M \\
CT & 4 & 192    & 897M \\ 
iCT & 4 & 192    & 897M \\ 
\midrule
\multicolumn{4}{l}{\textbf{Image Conditioned Model}} \\

HN-iCT-CN-Small & 2 & 32   & 14M \\ 
HN-iCT-CN-Mid & 4 & 32    & 22M \\  
\end{tabular}
\vspace{0.3cm} 
\caption{Model Configurations for unconditional image generation and image conditioned generation}
\label{tab:model_conf}
\end{table} 

\textbf{Training.}  We use the RAdam optimizer and learning rate is adjusted as 1-e4 for all datasets. For unconditional image generation, we train 800K iterations on CIFAR10 and CelebA 64x64 with a batch size of 512. 
For image conditional generation, we train 400K iterations on reconstructed CT slices with 512x512 patches with a batch size of 32.
 
\subsection{Unconditional Image Generation} 
We compare HN-iCT model with prior models are trained with CT, iCT technique and diffusion models. Considering lack of studies based on CT and iCT technique, we evaluate publicly available sources on iCT and CT technique for comparing with our approach. The models are gathered from public resources are referred as number of versions from $v1$ to $v3$ in table \ref{tab:unc_results}. Additionally, to recover fairness of comparison between models, all models  compared in table \ref{tab:unc_results} are trained locally. 
As shown in Table \ref{tab:unc_results}, HN-iCT performs better FID results when it is compared the other models which are publicly available and official models. We employed $2$ number of residual blocks instead of $4$ to prove robustness of our proposed training technique. 
\footnotetext[1]{iCT-v1: \url{https://github.com/Kinyugo/consistency_models}}
\footnotetext[2]{CT-v1:  \url{https://github.com/cloneofsimo/consistency_models}}
\footnotetext[3]{CT-v2: \url{https://github.com/junhsss/consistency-models}} 

\begin{table}[ht]
\centering
\begin{tabular}{lccccc}
\toprule
 & &  & \textbf{CIFAR10 32x32} & \textbf{CelebA 64x64}   \\
\cmidrule(lr){4-5}
\textbf{Model} & \textbf{\#Residual Blocks}  & \textbf{NFE} & \multicolumn{2}{c}{\textbf{FID $\downarrow$}} \\
\midrule
DDPM \cite{DDPM} & 4   & 1000 & 3.17 & 3.26   \\
DDIM \cite{DDIM} & 4  & 10 & 13.36 & 17.33   \\
EDM \cite{EDM} & 4   & 35 & 2.04 & 3.32   \\
CT \cite{Song2023-cm} & 4   & 1 & 14.32 & 18.72  \\
CT-v1\footnotemark[2] & 4   & 1 & 15.54 & 19.66   \\

iCT \cite{Song2023-imp} & 4   & 1 & 13.50 & 15.60   \\
iCT-v1\footnotemark[1] & 4  & 1 & 14.72 & 18.54   \\

iCT-v2\footnotemark[3] & 4  & 1 & 13.03 & 16.03   \\
\textbf{HN-iCT-Small} & 2   & 1 & \textbf{10.50} & \textbf{12.31}   \\
\bottomrule
\end{tabular}
\vspace{0.3cm} 
\caption{Model Performance on CIFAR10 32x32 and CelebA 64x64,  $\alpha=1.5$, $\beta=5$}
\label{tab:unc_results}
\end{table}
 
\subsection{Image Conditioned Generation} 
In this section, our proposed image conditioned consistency model (HN-iCT-CN Small) which employs Weighted Attention Gate (WAG) modules is utilized for enhancing low-dose CT scans. Experimental results are shown in Table \ref{tab:quantitative_results} and compared with related studies utilizing same LDCT dataset \cite{LDCTdataset}.

\begin{figure}[h]
    \centering
    \includegraphics[width=0.95\textwidth]{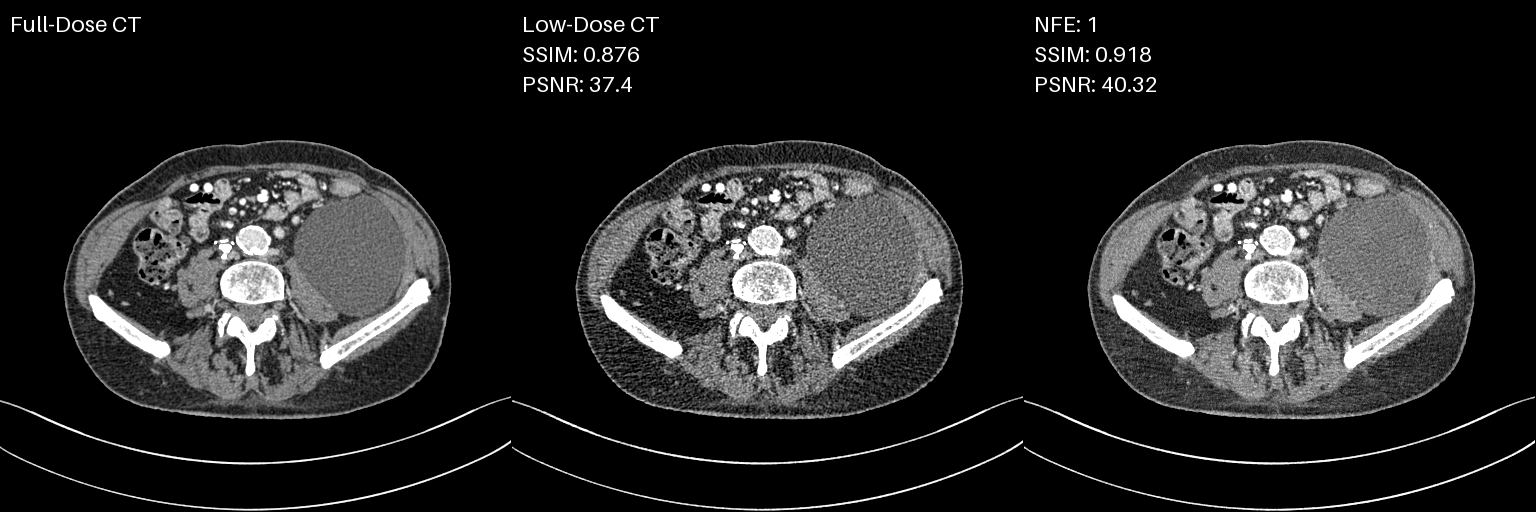}  
    \includegraphics[width=0.95\textwidth]{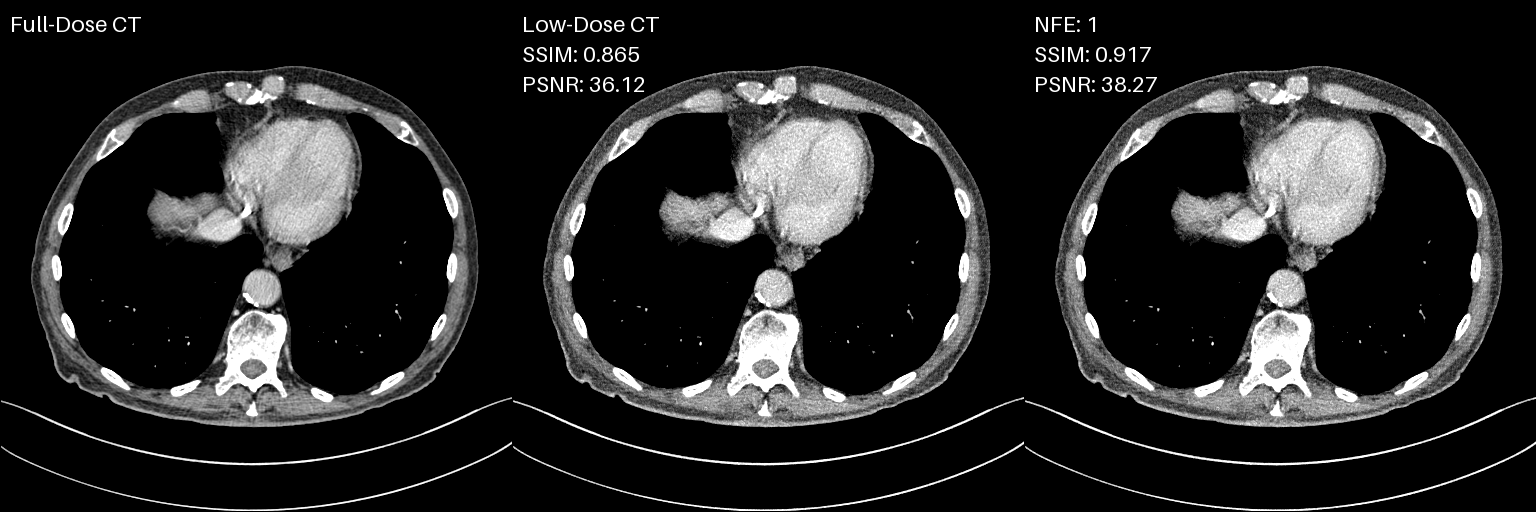}
    \includegraphics[width=0.95\textwidth]{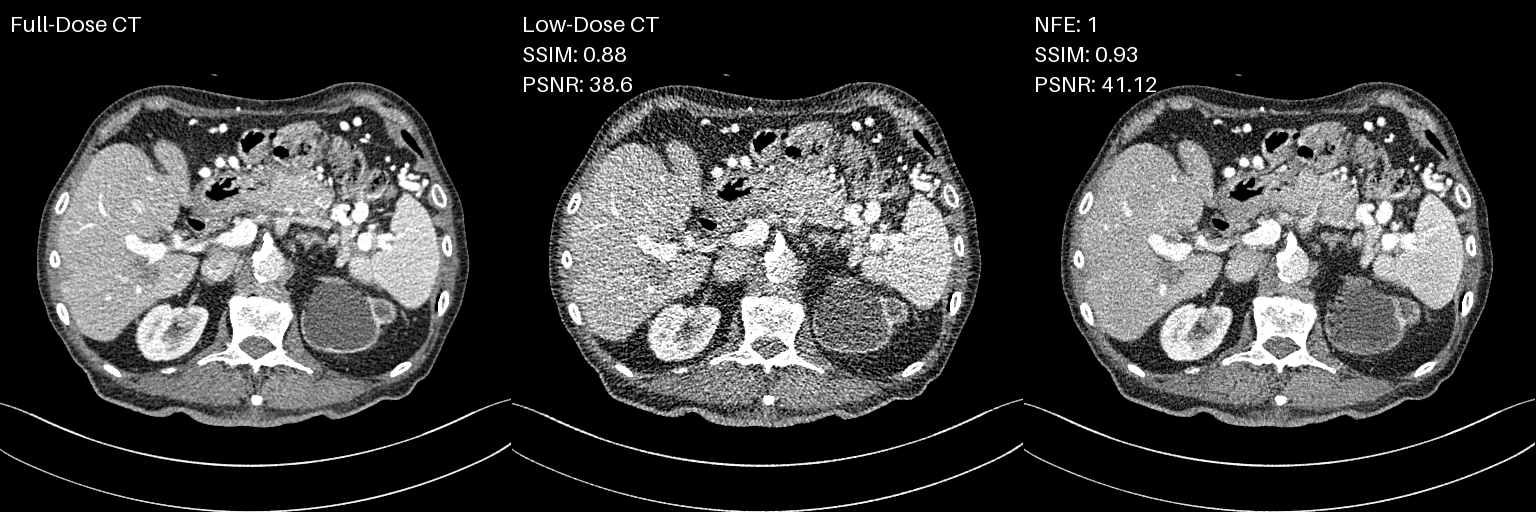}
    \caption{Representation of denoised slice samples from the low-dose CT validation set. The first column features full-dose slice samples, the middle column includes low-dose slices, and the last column presents single-step denoising samples processed by the HN iCT CN model. 1 mm slices, Window setting [-160,240] HU }
    \label{fig:ldct_samples}
\end{figure}

\begin{table}[h]
    \centering
    \renewcommand{\arraystretch}{1.5} 
    \begin{tabular}{ccccc}
        \hline
        \textbf{Model Name}& \textbf{LPIPS (↓)} & \textbf{SSIM (↑)} & \textbf{PSNR (↑)} & \textbf{NFE (↓)} \\
        \hline
        \textbf{LDCT} & $0.152 \pm 0.02$ & $0.94 \pm 0.02$ & $41.50 \pm 1.5$ &   \\
        \hline
        \textbf{DDPM \cite{DDPM_LDCT}} & $0.032 \pm 0.02$ & $0.97 \pm 0.02$ & $43.11 \pm 1.6$ & 50 \\
        \hline
        \textbf{EDM \cite{EDM}} &  $0.053 \pm 0.01$ &  $0.97 \pm 0.01$  & $43.80 \pm 1.2$ & 79 \\
        \hline
        \textbf{DDIM \cite{DDIM_LDCT} } & $0.053 \pm 0.01$ & $0.81 \pm 0.01$ & $35.35 \pm 1.2$ & 10 \\
        \hline
        \textbf{CD \cite{Song2023-cm}} &  $0.065 \pm 0.01$ &  $0.94 \pm 0.01$ & $42.00 \pm 0.8$ & 1 \\
        \hline 
        
        \textbf{PFGM++ \cite{PFGM++}} &  $0.055 \pm 0.01$ &  $0.96 \pm 0.01$ & $43.6 \pm 0.8$ & 79 \\
        \hline
        \textbf{PS-PFCM \cite{pfcm}} & $0.061 \pm 0.01$ & $0.96 \pm 0.01$ & $43.00 \pm 0.8$ & 1 \\
        \hline 
        \multicolumn{5}{c}{\textbf{HN iCT CN Small }} \\
        \hline
        \textbf{$\alpha$=0.5,$\beta$=5 } & $0.016 \pm 0.01$ & $0.96 \pm 0.01$ & $43.68 \pm 1.2$ & 1 \\ 
                \multicolumn{5}{c}{\textbf{HN iCT CN Medium }} \\
        \hline 
        \textbf{$\alpha$=0.5,$\beta$=5 } & $0.017 \pm 0.01$ & $0.96 \pm 0.01$ & $44.07 \pm 1.4$ & 1 \\ 
        
    \end{tabular}
    \vspace{0.5cm} 
    \caption{Quantitative Results, copied from other paper. }
    \label{tab:quantitative_results}
\end{table}
We explored that, effectiveness of the noise distribution on denoising performance depends and highly related on $\alpha$ parameter, to adjust the balance between low and high noise levels in generated noise levels. This reveals a correlation between batch size and the $\alpha$ parameter, where a reduction in batch size leads to a decrease in $\alpha$, directly impacting the overall performance.  Based on the results from different tests, the optimal $\alpha$ value for achieving the best denoising performance is set to 0.5. Additonally, experimental denoising samples from Low Dose CT validation set are represented in Figure \ref{fig:ldct_samples}. 

\subsection{Ablation Study} 
In this section, we present an ablation study conducted on the improved curriculum, sinusoidal curriculum, log-normal noise distribution, and beta noise distribution. The model configuration used for the ablations is designated as HN-iCT-Small, and the training steps are kept the same at 400K for each configuration, as shown in Table \ref{ablation_table}. 
Each training configuration for the ablations is represented by version numbers from $HN-iCT-S_{v1}$ to $HN-iCT-S_{v4}$, with two different options configured within the maximum possible transformation configurations.
\begin{table}[!ht]
\centering
\scalebox{0.9}{ 
\begin{tabular}{p{3cm}p{1.8cm}p{3cm}p{3.5cm}}
\hline
\textbf{Model Versions} &  \textbf{Cifar10} & \multicolumn{2}{c}{\textbf{Components}} \\  
\cline{3-4}
 & \textbf{FID} & \textbf{Curriculum Type} & \textbf{Noise Distribution} \\ 
\hline
$HN-iCT-S_{v1}$ & 21.19 & Improved & Log-Normal \\ 
\hline
$HN-iCT-S_{v2}$ & 17.31  & Improved & Beta \\ 
\hline
$HN-iCT-S_{v3}$ & 17.85 & Sinusoidal & Log-Normal \\ 
\hline
$HN-iCT-Small$ & \textbf{10.51} & Sinusoidal & Beta \\ 
\hline 
\end{tabular}}
    \vspace{0.3cm}
\caption{Ablation study on curriculum and noise distribution , evaluated at 400K training steps. $\beta$=1.5 , $\alpha$=5 for Beta distribution }
\label{ablation_table}
\end{table}
\section{Discussion and Conclusion}
Our enhancements to noise scheduling and curriculum address the need for a balanced noise distribution and controlled progression of noise steps within the curriculum. We examined the impact of the high noise levels in a noise distribution comprising the high-weighted low noise levels by beta distribution. Also, proposed curriculum which is based on sinusoidal function also provides a stability with minor changes on the number of noise steps for each training steps. Although the sinusoidal curriculum and beta distribution have a significant effect on the denoising performance of unconditional image generation, they need to be extended with different values for the parameters $\beta$, $\alpha$, $s_0$, and $s_1$. The biggest obstacle is widely known to be the training time. \\ 
In aspect of medical image processing, image conditioned architecture has made it possible to process medical images without resizing, by decreasing the number of learnable parameters around $80\%$. Also, it should be regarded that weighted Attention Gate modules have valuable impact on leveraging denoising performance for medical images. This experimental approach conducted on consistency models has a potential to combine different images given as condition and input,  and may lead  various applications by tuning parameters on WAG module. \\

In conclusion, it is presented a new curriculum schedule and noise distribution for consistency models, which provides flexibility on distribution of noise levels in a mini-batch by adjusting $\alpha$ and $\beta$ parameters. The suggested approach for unconditional image generation makes it possible to generate better quality images with lower model size and the model can learn easily trajectory from the prior noise distribution to posterior distribution of noise. The results indicate that tuning the weighting parameter in the WAG module, as well as the $\beta$ and $\alpha$ parameters in the curriculum and the beta noise distribution algorithm, allows us to outperform the corresponding consistency models. 

\section{Acknowledgements}
The project described was supported by the NIH National Center for Advancing Translational Sciences through grant number UL1TR001998. The content is solely the responsibility of the authors and does not necessarily represent the official views of the NIH.

\bibliographystyle{plain}
\bibliography{references.bib}

\begin{thebibliography}{10}

\bibitem{DDPM_Survey}
Florinel-Alin Croitoru, Vlad Hondru, Radu~Tudor Ionescu, and Mubarak Shah.
\newblock Diffusion models in vision: A survey.
\newblock {\em IEEE Transactions on Pattern Analysis and Machine Intelligence}, 45(9):10850–10869, September 2023.

\bibitem{DDPM_vs_GAN}
Prafulla Dhariwal and Alex Nichol.
\newblock Diffusion models beat gans on image synthesis, 2021.

\bibitem{BM3D}
P~Fumene~Feruglio, C~Vinegoni, J~Gros, A~Sbarbati, and R~Weissleder.
\newblock Block matching 3d random noise filtering for absorption optical projection tomography.
\newblock {\em Physics in Medicine and Biology}, 55(18):5401–5415, August 2010.

\bibitem{pfcm}
Dennis Hein, Adam Wang, and Ge~Wang.
\newblock Poisson flow consistency models for low-dose ct image denoising, 2024.

\bibitem{DDPM}
Jonathan Ho, Ajay Jain, and Pieter Abbeel.
\newblock Denoising diffusion probabilistic models, 2020.

\bibitem{Karras}
Tero Karras, Miika Aittala, Timo Aila, and Samuli Laine.
\newblock Elucidating the design space of diffusion-based generative models.
\newblock 2022.

\bibitem{EDM}
Tero Karras, Miika Aittala, Timo Aila, and Samuli Laine.
\newblock Elucidating the design space of diffusion-based generative models, 2022.

\bibitem{cifar10}
Alex Krizhevsky.
\newblock Learning multiple layers of features from tiny images.
\newblock pages 32--33, 2009.

\bibitem{mean_filter}
Zhoubo Li, Lifeng Yu, Joshua~D. Trzasko, David~S. Lake, Daniel~J. Blezek, Joel~G. Fletcher, Cynthia~H. McCollough, and Armando Manduca.
\newblock Adaptive nonlocal means filtering based on local noise level for ct denoising: Adaptive nonlocal means filtering for ct denoising.
\newblock {\em Medical Physics}, 41(1):011908, December 2013.

\bibitem{DDIM_LDCT}
Xuan Liu, Yaoqin Xie, Jun Cheng, Songhui Diao, Shan Tan, and Xiaokun Liang.
\newblock Diffusion probabilistic priors for zero-shot low-dose ct image denoising, 2023.

\bibitem{celeba}
Ziwei Liu, Ping Luo, Xiaogang Wang, and Xiaoou Tang.
\newblock Deep learning face attributes in the wild.
\newblock In {\em Proceedings of International Conference on Computer Vision (ICCV)}, December 2015.

\bibitem{LDCTdataset}
Cynthia~H. McCollough, Adam~C. Bartley, Rickey~E. Carter, Baiyu Chen, Tammy~A. Drees, Phillip Edwards, David~R. Holmes, Alice~E. Huang, Farhana Khan, Shuai Leng, Kyle~L. McMillan, Gregory~J. Michalak, Kristina~M. Nunez, Lifeng Yu, and Joel~G. Fletcher.
\newblock Low‐dose <scp>ct</scp> for the detection and classification of metastatic liver lesions: Results of the 2016 low dose <scp>ct</scp> grand challenge.
\newblock {\em Medical Physics}, 44(10), October 2017.

\bibitem{attention_U_NET}
Ozan Oktay, Jo~Schlemper, Loic~Le Folgoc, Matthew Lee, Mattias Heinrich, Kazunari Misawa, Kensaku Mori, Steven McDonagh, Nils~Y Hammerla, Bernhard Kainz, Ben Glocker, and Daniel Rueckert.
\newblock Attention u-net: Learning where to look for the pancreas, 2018.

\bibitem{DDIM}
Jiaming Song, Chenlin Meng, and Stefano Ermon.
\newblock Denoising diffusion implicit models, 2020.

\bibitem{Song2023-imp}
Yang Song and Prafulla Dhariwal.
\newblock Improved techniques for training consistency models.
\newblock 2023.

\bibitem{Song2023-cm}
Yang Song, Prafulla Dhariwal, Mark Chen, and Ilya Sutskever.
\newblock Consistency models.
\newblock 2023.

\bibitem{score_based}
Yang Song, Jascha Sohl-Dickstein, Diederik~P Kingma, Abhishek Kumar, Stefano Ermon, and Ben Poole.
\newblock Score-based generative modeling through stochastic differential equations.
\newblock 2020.

\bibitem{butterfly}
Josiah Wang, Katja Markert, and Mark Everingham.
\newblock Learning models for object recognition from natural language descriptions.
\newblock In {\em Proceedings of the British Machine Vision Conference}, 2009.

\bibitem{DDPM_LDCT}
Wenjun Xia, Qing Lyu, and Ge~Wang.
\newblock Low-dose ct using denoising diffusion probabilistic model for 20$\times$ speedup, 2022.

\bibitem{PFGM++}
Yilun Xu, Ziming Liu, Yonglong Tian, Shangyuan Tong, Max Tegmark, and Tommi Jaakkola.
\newblock Pfgm++: Unlocking the potential of physics-inspired generative models, 2023.

\bibitem{GAN_survey}
Xin Yi, Ekta Walia, and Paul Babyn.
\newblock Generative adversarial network in medical imaging: A review.
\newblock {\em Medical Image Analysis}, 58:101552, December 2019.

\bibitem{beta_diff}
Mingyuan Zhou, Tianqi Chen, Zhendong Wang, and Huangjie Zheng.
\newblock Beta diffusion, 2023.

\end{thebibliography}

\begin{appendices} 
\section{High Noise Level Experimental Details} \label{appendix_A}

The experiments reveals that adding minor weighted high noise levels on mini-batches increase denoising performance. As it is represented in table \ref{tab:table_high_noise}, while adding high level noise levels with lower percentage ratios can enhance denoising performance, adding high level noise at $10\%$ of the mini-batch length has effects on denoising performance conversely.

\begin{table*}[ht]
\scriptsize
\centering
\caption{Evaluation of denoising performance of models trained with noise schedules, including high noise levels proportional to the size of mini-batches.}
\resizebox{\linewidth}{!}{%
\begin{tabular}{l cccccccccc} 
\toprule
Model Versions  & Batch Size & Training Steps & Number of Res. Blocks     & High Level Noise Ratio  &N & Noise Scheduling & Attention Resolutions  & FID Score \\   
\midrule

iCT v1 & 1024 & 100000 & 2    & 0\% &100  & Log-Normal & [16,8]        & 66.51  \\ 
\midrule

iCT v2 & 1024 & 100000 & 2    & 2\% &100  & Log-Normal & [16,8]        & 75.34  \\ 
\midrule
iCT v3  & 1024 & 100000  & 2      & 3\% &100  & Log-Normal & [16,8]    & 43.81  \\  
\midrule
iCT v4  & 1024 & 100000  & 2      & 4\% &100  & Log-Normal & [16,8]     & 40.32  \\   
\bottomrule
iCT v5  & 1024 & 100000  & 2      & 5\% &100  & Log-Normal & [16,8]     & 73.24  \\   
\bottomrule
\end{tabular}%
}
\label{tab:table_high_noise}
\end{table*}
The results represented in Table \ref{tab:table_high_noise} demonstrate that implementing a log-normal noise distribution with high noise levels enhances denoising performance when the high-level noises reach $5\%$ of the total number of mini-batch size. Based on these results, we propose beta noise distribution  which represents high-weighted low noise levels and low-weighted high noise levels. Additionally, beta disribution provides a flexibility to adjust on the ratio of noise levels by adjusting $\beta$ and $\alpha$ parameters as it desired. 

\end{appendices}

\end{document}